\documentstyle[aps,epsf,twocolumn]{revtex}

\def\la{\langle}
\def\ra{\rangle}

\def\Tr{{\,\rm Tr}}

\def\Lsu2{{\cal L}_{\mbox{SU(2)}}}

\def\su2xsu2{{SU(2)\times SU(2)}}
\def\su3xsu3{{SU(3)\times SU(3)}}

\def\be{\begin{eqnarray}}
\def\ee{\end{eqnarray}}
\def\ben{\begin{eqnarray}}
\def\een{\end{eqnarray}}

\newcommand{\beq}{\begin{equation}}
\newcommand{\eeq}{\end{equation}}
\newcommand{\bea}{\begin{eqnarray}}
\newcommand{\eea}{\end{eqnarray}}
\newcommand{\tr}{{\rm tr}}
\newcommand{\no}{\nonumber}
\begin{document}
\title{\bf  Dirac Spectrum in QCD and Quark Masses}
\author{{\bf Jerzy Jurkiewicz}$^{1,2}$, {\bf Maciej A. Nowak}$^{2,3}$  
and {\bf Ismail Zahed}$^4$}

\address{
$^1$LPTHE, Universit\'{e} de Paris XI, B\^{a}timent 211,  91405 Orsay Cedex,
 France;\\
$^2$ Department of Physics, Jagellonian University, 30-059 Krakow, 
Poland;\\
$^3$ GSI, Plankstr. 1, D-64291 Darmstadt, Germany \& \\
 Institut f\"{u}r Kernphysik, TH Darmstadt, D-64289 Darmstadt, Germany \\
$^4$Department of Physics, SUNY, Stony Brook, New York 11794, USA.}

\maketitle
                                                                               
\begin{abstract}
We use a chiral random matrix model to investigate the effects of massive 
quarks on the distribution of eigenvalues of QCD inspired Dirac operators.
Kalkreuter's lattice analysis of the spectrum of the massive (hermitean)
Dirac operator for two colors and Wilson fermions is shown to follow from
a cubic equation in the quenched approximation. The quenched spectrum shows a  
Mott-transition from a (delocalized) Goldstone phase softly broken by the 
current mass, to a (localized) heavy quark phase, with quarks localized over 
their Compton wavelength. Both phases 
are distinguishable by the quark density of states at zero virtuality, with
a critical quark mass of the order of 100-200 MeV. At the critical point, the
quark density of states is given by $\nu_Q (\lambda) \sim |\lambda|^{1/3}$.
Using Grassmannian techniques, we derive an integral representation for the 
resolvent of the 
massive Dirac operator with one-flavor in the unquenched approximation, 
and show that near zero virtuality, the distribution of eigenvalues is 
quantitatively changed by a non-zero quark mass. The generalization of our 
construction to arbitrary flavors is also discussed. Some recommendations for 
lattice simulations are suggested.

\end{abstract}
\vskip 0.6cm

{\bf 1. \,\,\, Introduction}\\

In QCD, the character of the Dirac spectrum near zero virtuality follows from 
symmetries alone, in the limit where the quark mass is taken to zero. 
This is best exemplified by the Banks-Casher relation \cite{BANKS}
for the quark density of states, and its moments  \cite{SMILGA}. A number of 
recent studies have confirmed these relations using a chiral random matrix 
formulation, and unraveled their generic structure around zero 
virtuality (microscopic or mesoscopic limit) in terms of a universal 
spectral density \cite{JAC,JACMANY}.

Given the importance of the concept of current quark masses for spontaneous 
symmetry breaking and restoration in QCD, an important yet unanswered 
question in the context of chiral random matrix models, has 
to do with the role of a finite current quark mass. The $1/N$ expansion used
in the Coulomb gas approach calls for non-trivial subleading effects 
\cite{MEHTA,MIGDAL}. The method of orthogonal polynomials \cite{MEHTA}
fails in the massive case \cite{JAC}. Recently, Kazakov \cite{KAZAKOV}, 
Brezin, Hikami and Zee (BHZ) \cite{BREZIN}, and Zee \cite{ZEE} have discussed
a number of alternative and powerful methods to analyze the spectral densities 
and level correlations of a large class of 
random matrix models.
 Their methods borrow from exact integral representations, 
Grassmannian 
formulations and diagrammatic techniques. Some of these methods will
be taken up in this paper and applied to various chiral random matrix models 
as inspired by QCD spin and flavor symmetries. 
In section 2, we go over the general aspects of the 
spectral 
density, its relation to the quark condensate and its analogy with the 
Kubo formula. The striking violation of Lifshitz's bound \cite{LIF}
near zero virtuality is also discussed. In section 3, we recall results from 
chiral random matrix models 
both in the microscopic and macroscopic limit. In section 4, we analyze 
Kalkreuter's quenched lattice SU(2) simulations of the spectral density 
using recently developed arguments by Zee \cite{ZEE} and Kazakov \cite{KAZAKOV}.
We show that for a current mass $ m = -(N/V_4) \la \overline q q \ra^{-1}$
(where $N/V_4$ is the number of quark states in the four Euclidean volume,
and $\la \overline{q} q \ra$ the quark condensate),  
the spectral  distribution for the unsquared and hermitean Dirac operator
for Wilson fermions, shows
 a phase transition from a Goldstone phase to a phase where the quarks are 
localized over their Compton wavelength, with strong chiral symmetry breaking.
The order parameter in this case is the quark density of states at zero 
virtuality, and the transition is reminiscent of a conductor-insulator (Mott) 
transition. In section 5, the unquenched problem is analyzed using
the 
Grassmaniann method introduced by BHZ. A general integral representation for 
the spectral density is derived. In section 6, we show that the integral 
representation leads to the known spectral density in the macroscopic as well 
as microscopic limit for $N_F=1$ and zero quark mass. In section 7, the case
of non-zero quark mass is investigated, leading to a new spectral density
in the microscopic limit and for quark masses $mN\sim 1$. The generalization 
to an arbitrary number of flavors is outlined. Our conclusions and 
recommendations are summarized in section 8.

\vskip .6cm
{\bf 2.\,\,\, QCD Spectral Distribution}\\

The spectral representation for the quark propagator of flavor
$F$ in a fixed gluon background is given by
\be
S_F(x,y,A)= \sum \frac{\phi_n(x) \phi_n^{\dagger}(y)}{-\lambda_n-im_F}
\ee
where $\phi_n, \,\lambda_n$ are eigenvectors and eigenvalues of the massless
Dirac equation
\be
i\rlap/\nabla(A)\phi_n(x) = \lambda_n \phi_n(x).\label{eigen}
\ee
The fermion condensate in Euclidean space is 
\ben
\la q^\dagger q\ra  = -\sum_{F=1}^{N_F}
\la\la {\Tr} S_F(x,x, A)\ra\ra \label{eferm}
\een
where $\la\la \dots \ra\ra $ denotes the averaging over the gluonic 
configurations $A$ using the QCD action with massive quarks. In the limit 
where the four-volume $V_4$ goes to infinity, the  spectrum becomes dense 
and we may use the eigenvalue density (\ref{eigen})
\be
\nu (\lambda , m_F ) = \la\la \sum_n \delta (\lambda -\lambda_n )\ra\ra.
\label{meanspec}
\ee
to characterize the Dirac spectrum.
Through the fermion determinant in the averaging measure, 
(\ref{meanspec}) carries a nontrivial dependence on the current masses $m_F$.
We will refer to the mass dependence in (\ref{meanspec}) as the sea mass
dependence.
In terms of (\ref{meanspec}) the fermion condensate (\ref{eferm}) becomes
\be
\la q^{\dagger}q\ra  
   = \frac{1}{V_4}\sum_{F=1}^{N_F}
\int d\lambda \frac{\nu(\lambda , m_F )}{\lambda +im_F}.
\label{cond1}
\ee
The explicit mass dependence in the denominator of (\ref{cond1}) will be 
referred to as the valence mass dependence, for obvious reasons. In QCD, 
both the sea and valence masses are the same. Here, we may choose to 
disentangle them (for theoretical clarity) whenever indicated. Throughout, 
we will think of the masses as fixed external parameters, although in QCD
the quantum averaging forces them to run. This brings about the nasty issue
of the ultra-violet sensitivity of (\ref{cond1}) for finite current quark 
masses.
Although perturbative renormalization of (\ref{cond1}) is possible we will
not discuss it here. Most of the discussions to follow,
 focuses on the infrared 
part of the spectrum (around zero virtuality). The random matrix models 
to be discussed below are inspired models for the constant quark modes only,
with some astonishing resemblance to lattice regulated simulations.

In the chiral limit, $m_F \rightarrow 0$, the Dirac operator $i\rlap/\nabla$
anticommutes with $\gamma_5$, so the non-zero eigenvalues come in pairs 
$(\lambda, -\lambda)$ and the spectral function is symmetric. Thus
\be
\la \bar{q}q\ra = - \frac{\pi N_F}{V_4} \nu(0).
\label{banks_casher}
\ee
following a Wick rotation to Minkowski space $(q^{\dagger} , q )\rightarrow 
(i\overline{q}, q)$.  
This relation was first derived by Banks and Casher\cite{BANKS}.
It is important that the chiral limit is sampled with a valence 
quark mass $m_F$ that is taken to zero after 
the thermodynamical limit $V_4\rightarrow \infty$, for otherwise the result
would be zero. The spontaneous breakdown of a continuous symmetry cannot take 
place in finite volumes, unless the condition $m_F\la \overline q q\ra V_4 >>1$
is fulfilled. The result (\ref{banks_casher}) states 
that in vector-like theories with chiral symmetric (even)
spectra, the quark condensate
is related to the mean spectral density at zero virtuality ($\lambda =0$).
The delocalization of the quark modes is 
caused by strong correlations that randomize the Dirac spectrum near zero,
triggering a huge accumulation at zero. As $V_4\rightarrow\infty$ the number 
of eigenvalues grows with the four volume $V_4$ in contrast to the length
$^4\sqrt{V_4}$ in the free  case\cite{SMILGA,CASHER}. 

The change in the 
number of quark states near zero virtuality is drastically different from 
what is expected from Lifshitz's condition \cite{LIF,NEU} for the case of 
scattering off random repulsive centers, where the density of states is 
found to vanish exponentially. The reason may be traced back to the 
chirality structure of the random ensemble discussed here, hence to the spin
of the quarks in QCD. 
Lifshitz's condition can be evaded by noting that in a magnetic field 
further delocalization can be generated without cost of energy \cite{CASHER}.
For spinless particles the density of states is bounded from above 
by the free quark density of states \cite{BANKS}, hence no condensate
is allowed to form in scalar QCD.

The Banks-Casher relation (\ref{banks_casher}) is reminiscent
of the conductivity in metals, where the latter is proportional to the density 
of states at the Fermi surface. Indeed, it follows from the Kubo formula that 
the d.c. conductivity $\sigma$ relates to the density of states
at the Fermi level, $\rho ( E_F)$,  through
\be
\sigma = e^2 \,\,{\bf D}\,\, \rho (E_F)
\label{kubo}
\ee
where $e$ is the electron charge and ${\bf D}$ the diffusion constant.
This result (\ref{kubo}) is reminiscent of (\ref{banks_casher})
with the 
identification of  $\pi\sigma/e^2{\bf D}$ with $-\la \overline q q\ra$,
and $E_F$ with $\lambda\sim 0$.

\vskip .6cm
{\bf 3.\,\,\, Random Matrix Model}\\

The pertinent random matrix model for the QCD spectrum near zero virtuality 
follows from the color representation of the quark fields (here fundamental), 
the chirality odd character of the massless Dirac operator in a fixed gluon 
background and the number of flavors.  In the continuum, 
the QCD Dirac operator 
for three colors may be mapped onto the Gaussian Unitary Ensemble (GUE), while
for two colors it may be mapped onto the Gaussian Othogonal Ensemble (GOE)
\cite{JACMANY}. 
In the zero topological charge sector, and for three colors the generating 
functional is \cite{JAC,JACMANY,NOWAK,SEMENOV,SHURYAK},
\be
Z [m] = \int \,\, dT\,\, \prod_{F}^{N_F}\,\,{\rm det}
\left(\matrix{im_F & T\cr T^{\dagger} & im_F\cr}\right)\,\,e^{-N\Sigma^2{\Tr} 
(T^{\dagger} T )}
\label{RM1}
\ee
where $T$ is a random $N\times N$ complex matrix, with $N$ identified with 
the four volume $V_4$ and $\Sigma$ the chiral condensate appearing in the 
Banks-Casher relation (\ref{banks_casher}).
Equation (\ref{RM1}) is the generating function for the GUE.
The joint eigenvalue density following from (\ref{RM1}) reads\cite{JAC}
\be
\nu (\lambda_1, ..., \lambda_N; m_F ) = &&{\bf C}_N 
\prod_{i\leq j} |\lambda_i^2 -\lambda_j^2 |^2\nonumber\\
&&\prod_i (\lambda_i^2 + m_F^2)^{N_F} |\lambda_i|
\, e^{-N\Sigma^2 \sum_{l=1}^N \lambda_l^2}.
\label{RM2}
\ee
with ${\bf C}_N$ an overall normalization (see below).
Integrating (\ref{RM2}) over $(N-1)$-eigenvalues and taking the 
microscopic limit $N\rightarrow\infty$ with $x=N\lambda$ fixed,\footnote{Note 
that our definition differs slightly from the one used in 
reference~\cite{JAC} due to replacement
$2N \leftrightarrow N$, therefore $\nu_s(x) \leftrightarrow 2 \nu_s(2x)$.}  
yields the following form for the microscopic
spectral density ($m_F=0$) \cite{JAC} 
\be
\nu_s (x) = 2\Sigma^2 x  \bigg( J_{N_F}^2 (2\Sigma x ) -
J_{N_F-1} (2\Sigma x )J_{N_F+1} (2\Sigma x )\bigg)
\label{RM3}
\ee
which is the master formula for all the sum rules discussed by Leutwyler and 
Smilga\cite{SMILGA}.
  Equation  (\ref{RM3}) shows that around zero
virtuality ($\lambda \sim 0$), the distribution of eigenvalues 
oscillates to zero. For $N_F=0$ these oscillations are caused by the level 
repulsion around zero due to the symmetric character of the spectrum under
chirality (Airy phenomenon \cite{BREZIN}). These oscillations are affected 
by the fermionic determinant in the massless case, as is clear from the 
$N_F$ dependence (zero mode suppression). In the quenched approximation, the 
character of these oscillations appear to be unaffected by the choice of the
measure, provided that it is local. Non-local changes to the measure, $e.g.$
through a fermion determinant, do affect the structure of these oscillations.
Other non-local changes are also possible, but will not be discussed in this
work.

The joint eigenvalue density associated to (\ref{RM2}) can be obtained through 
a direct integration, or by using a Coulomb gas description of (\ref{RM1}). 
Indeed, using the definition 
(\ref{meanspec}) for $\nu ({\lambda})$, where the average is over the joint 
eigenvalue density (\ref{RM2}), allows a rewriting of (\ref{RM1}) in terms of 
an effective action
\be
{\bf S} [\nu ] =&& -\int \, d\lambda d\lambda' \,\nu(\lambda ) \nu(\lambda ')\,
{\rm ln} |\lambda^2-\lambda'^2|\nonumber\\
&&-\int \, d\nu \, \nu (\lambda ) \left( \sum_{F=1}^{N_F} 
{\rm ln} (\lambda^2 + m_F^2) + {\rm ln} |\lambda| \right)\nonumber\\
&&+N\Sigma^2 \int \, \nu (\lambda ) \, \lambda^2\nonumber\\
&& + \xi \left( \int \, d\lambda \, \nu (\lambda ) -N\right)\nonumber\\
\label{AR2}
\ee
where $\xi$ is a Lagrange multiplier. For a dense spectrum, the integration 
over the eigenvalues $\lambda_i$ may be traded by  a functional integration 
over the eigenvalue density $\nu (\lambda)$, modulo a Jacobian \cite{MIGDAL}. 
In the large $N$ limit, the extremum of ${\bf S} [\nu]$ including the 
contribution from the Jacobian to the effective action,
determines the macroscopic spectral density. Variation of 
(\ref{AR2}) with respect to $\nu$ and differentiation with respect to 
$\lambda$, yield
\be
2{\bf P} \int \frac{\nu (\lambda')}{\lambda^2 -\lambda'^2}
=N\Sigma^2 - \sum_{F=1}^{N_F} \frac{1}{\lambda^2+m_F^2} -\frac 
1{2\lambda^2} + {\bf J}
\label{AR3}
\ee
with the normalization condition
\be
\int d\lambda\,\, \nu (\lambda ) = N
\label{AR4}
\ee
The contribution ${\bf J}$ is due to the Jacobian and is of order $1/N^2$
\cite{MIGDAL}. Its explicit form will not be needed here. 
In (\ref{AR3}), ${\bf P}$ stands for the principal value of the integral.
In the thermodynamical limit ($N\rightarrow\infty$), the fermions and the 
Jacobian drop from the macroscopic spectral density in the chiral limit. Thus
\be
{\bf P} \int \frac{\nu (\lambda')}{\lambda^2 -\lambda'^2}
=N\frac{\Sigma^2}2
\label{AR5}
\ee
This integral equation yields a semi-circular distribution for the macroscopic 
spectral density
\be
\nu ({\lambda}) = \frac {N\Sigma}{\pi}\sqrt{1-\frac{\lambda^2 \Sigma^2}4}
\label{AR6}
\ee
The level 
repulsion revealed in the microscopic limit takes place in a window of 
size $1/N$ around the origin, and shrinks to zero size in the thermodynamical 
limit. 
In terms of (\ref{AR6}), the quark condensate is $\la \overline{q} q \ra 
= -N_F (N/V_4) \Sigma$. Here 
 $N/V_4$ is just the eigenvalue density of the Dirac 
operator in Euclidean space.
Throughout $\Sigma$ and $V_4$ are set to 
one, and the thermodynamical limit is understood for $N\rightarrow\infty$. 
The scale $\Sigma$ can be reinstated at the end by inspection.
Since the thermodynamical limits are now understood, we prefer to work 
from this moment with the macroscopic spectral density (\ref{AR6}) 
normalized to one instead of $N$, {\it i.e.},
\be
\nu(\lambda)=\frac{1}{2\pi} \sqrt{4-\lambda^2}.
\label{newnorm}
\ee
unless specified otherwise. 

\vskip .6cm
{\bf 4.\,\,\, Quenched Spectral Distributions}\\

Recent detailed numerical analysis  by Kalkreuter 
\cite{REUTER} using Wilson as well as staggered fermions,
provides some useful insights to the macroscopic character of the
fermionic spectrum of four-dimensional  gauge theories.
On the lattice, the spin-Lorentz structure of the QCD Dirac operator is 
affected. Indeed, for any number of colors greater than two\footnote{For two
colors the ensemble is the Gaussian Orthogonal Ensemble (GOE).}
 and Wilson fermions the 
pertinent random matrix ensemble is the Gaussian Unitary Ensemble (GUE).
The chiral structure is upset by Wilson's r-terms, needed to remove the
lattice doublers. For two colors and staggered fermions the chirality 
structure is preserved, but the Lorentz structure is upset. The pertinent 
random matrix ensemble is the Gaussian Symplectic Ensemble (GSE)
\cite{JACMANY}. For three colors it is back to the GUE.

\vskip .5cm
{$\bullet\,\,\,\,$ Wilson Fermions}
\vskip .2cm

In the quenched approximation, Kalkreuter's results for $SU(2)_c$ 
with Wilson fermions
for the unsquared and unnormalized operator 
${\bf Q}= \gamma_5 (\rlap/{\nabla} + m )$ 
(continuum), may be mocked up by the Gaussian Orthogonal Ensemble (GOE), 
provided that the gauge configurations are sufficiently random. Specifically
\footnote{For one flavor we set $m_F=m$.},
\be
{\bf Q}_W = \left(\matrix{m & 0\cr 0 & -m \cr}\right) + {\bf R}
\label{Kwil}
\ee
which is the sum  of a deterministic $2N\times 2N$ matrix (first
contribution, with $m$ - diagonal block $N\times N$),
and a random hermitean (symmetric) $2N\times 2N$ matrix. 
We note that ${\bf Q}_W^{\dagger}=
{\bf Q}_W$. The measure is
\be
{\bf P} ({\bf R}) = \frac 1Z \,\,e^{- 2N {\rm Tr} ({\bf R}{\bf R}^{\dagger} )} 
\label{KK2}
\ee
This problem is reminiscent of an electron scattering on impurities in a spin 
dependent quantum Hall fluid, as recently suggested by Zee \cite{ZEE}. 

The distribution of eigenvalues of ${\bf Q}_W$
\be
\nu_Q (\lambda , m ) = \frac{1}{2N} <{\rm Tr}_{2N} \,\delta\, 
(\lambda - {\bf Q}_W )>
\label{KK3}
\ee
follows from the resolvent ${\bf G} (z, m)$ of ${\bf Q}_W$ through
\be
\nu_Q (\lambda , m ) = -\frac 1{\pi} {\rm Im} \,\,{\bf G} (\lambda + i0 , m )
\label{KK4}
\ee
Since ${\bf R}$ is only hermitean (symmetric),
 the distribution of eigenvalues is only 
symmetric about zero virtuality on the average. 
In the chiral limit, the states 
are not necessarily paired about zero, because of the r-terms in the Wilson 
action.

The problem of determining the  one-point Green function ${\bf G} (z )$
for the sum of a deterministic 
Hamiltonian $H_0$ with eigenvalues $\epsilon_i,(\,i=1,\ldots,N)$ 
and a Gaussian random matrix was first solved by Pastur\cite{PASTUR},
and recently rederived and generalized using much simpler arguments by
Zee \cite{ZEE}. Generally, the Green function  is determined from  
\be
{\bf G}(z ) =\frac{1}{N} \sum_{i=1}^N 
\frac{1}{z-{\epsilon}_i - {\bf G }(z )}.
\label{pastureq}
\ee
For the case (\ref{Kwil}), the deterministic hamiltonian is composed of 
two diagonal blocks with eigenvalues $\pm m$, so 
${\bf G} (z , m )$ satisfies the cubic equation
\be
{\bf G}^3 - 2 z {\bf G}^2 + (z^2-m^2+1) {\bf G} - z =0
\label{KK5}
\ee
Using Cardano's complex  solution to (\ref{KK5}) we get 
\be
\nu_Q (\lambda , m  ) =\frac{\sqrt{3}}{2}\left[(r+\sqrt{d})^{1/3}-
(r-\sqrt{d})^{1/3}\right]
\label{KK6}
\ee
with 
\be 
r&=&\frac{1}{6}\lambda \left(1+2m^2-\frac{2}{9}\lambda^2\right) \nonumber \\ 
d&=&\frac{1}{27}\left[ (1-m^2)^3-\lambda^2(\frac{1}{4}-5m^2-2m^4+m^2\lambda^2)
\right]
\label{carcoeff}
\ee
At the critical point $m_*\sim 1$ and near zero virtuality $\lambda\sim 0$, 
the order parameter behaves as $\nu_Q (\lambda , 1) \sim |\lambda |^{1/3}$, 
with the critical exponent $\beta=1/3$. This is to be contrasted with the 
density of states of electrons near the mobility edge, rescattering off
random impurities and undergoing Anderson localization with $\beta =0$ 
\cite{WEGNER,ABRA,STONE}. 

It is worth mentioning, that although the random matrix model under 
consideration is only valid for the constant quark modes, it does reproduce
the bulk characteristic of the spectral function from lattice regulated 
calculations. This justifies $a$ $posteriori$ our assumption in ignoring the 
ultraviolet aspects of the quark spectrum, with their inherent diverging 
contribution to the quark condensate. We suspect that a cooling of Kalkreuter's
gauge
 configurations will not affect considerably the character of the spectral 
distribution. Such procedure can be used to define unambiguously the quark
condensate for finite current masses.

The distribution of eigenvalues of ${\bf Q}_W$
is strikingly similar to the one discussed recently in the finite temperature 
problem for the hermitean and massless Dirac operator
$i\rlap/\nabla$ on the torus in the quenched approximation 
\cite{Stephanov,Jackson,Jacksonplus}. This may be understood if we note 
that in the
high temperature phase, dimensional reduction implies that in Euclidean space
and in one-dimension lower, the effective quark mass is not $m$ but 
$\sqrt{m^2+\pi^2 T^2}$ after suitable chiral rotations \cite{Hansson}. In three
dimensions the random ensemble is indeed the unitary ensemble 
\cite{JACZAHED}. This observation can be used to map out the chiral sensitive 
part of the QCD phase diagram with Wilson fermions in the plane spanned
 by the temperature and current quark mass. We note 
that in the context of phase transitions, and if ${\bf G}$ (or at least its 
discontinuity along the real axis) is understood as an order parameter, then
the cubic equation (\ref{KK5}) is generic of second order phase transitions, as
expected from universality arguments. Some of these issues will be discussed 
elsewhere.

\begin{figure}[tbp]
\centerline{\epsfxsize=8.6cm \epsfbox{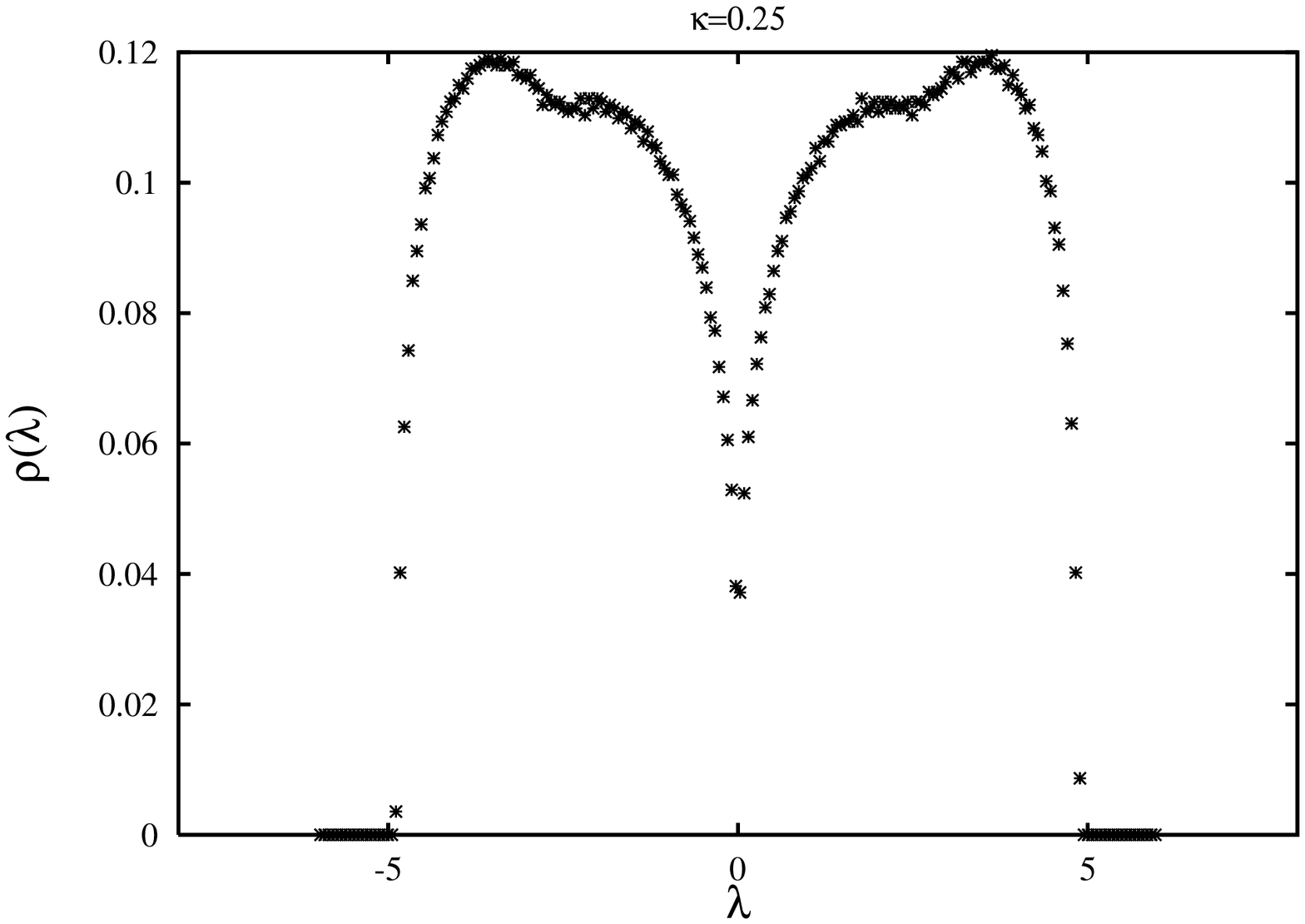}}
\centerline{\epsfxsize=8.6cm \epsfbox{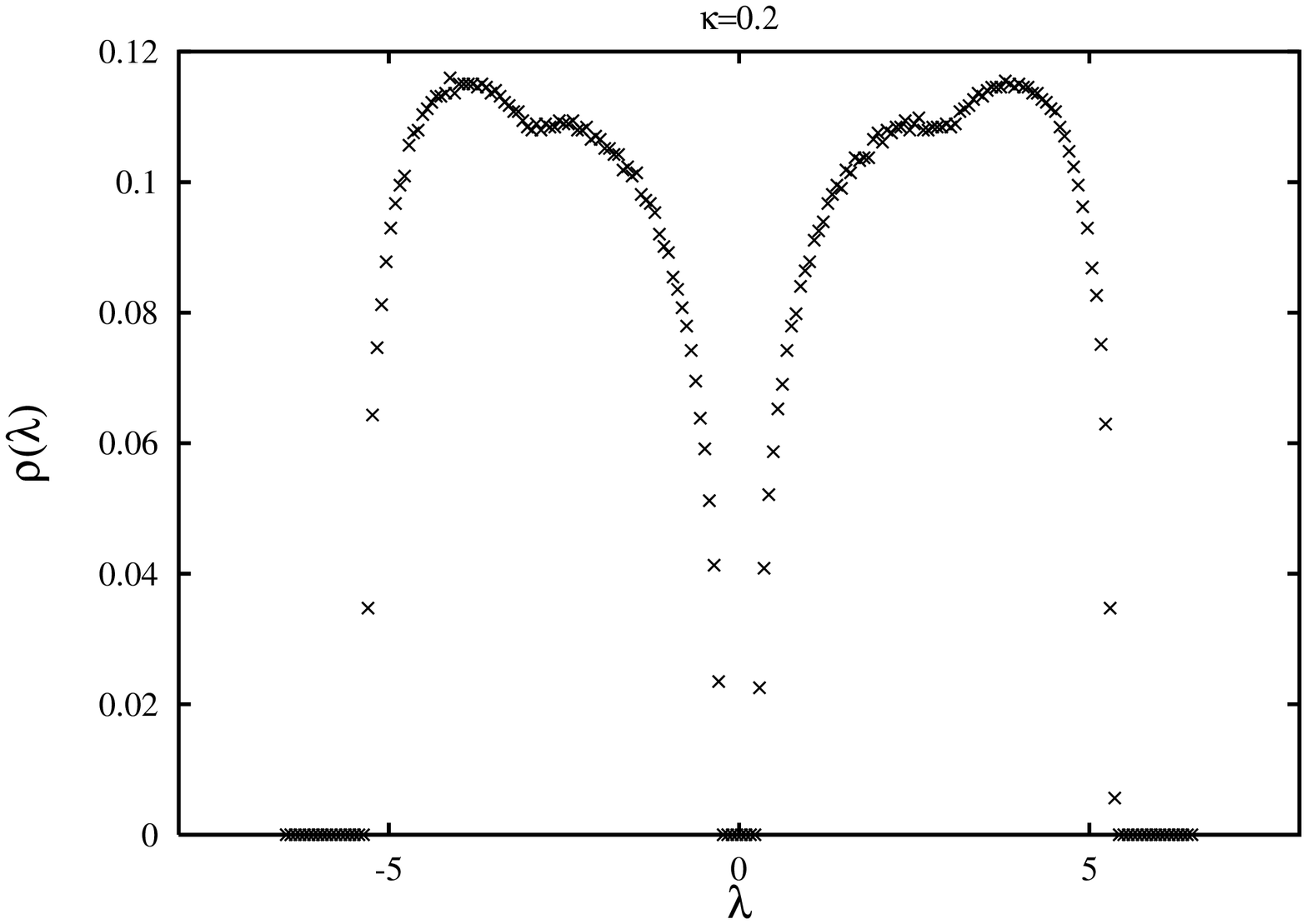}}
\centerline{\epsfxsize=8.6cm \epsfbox{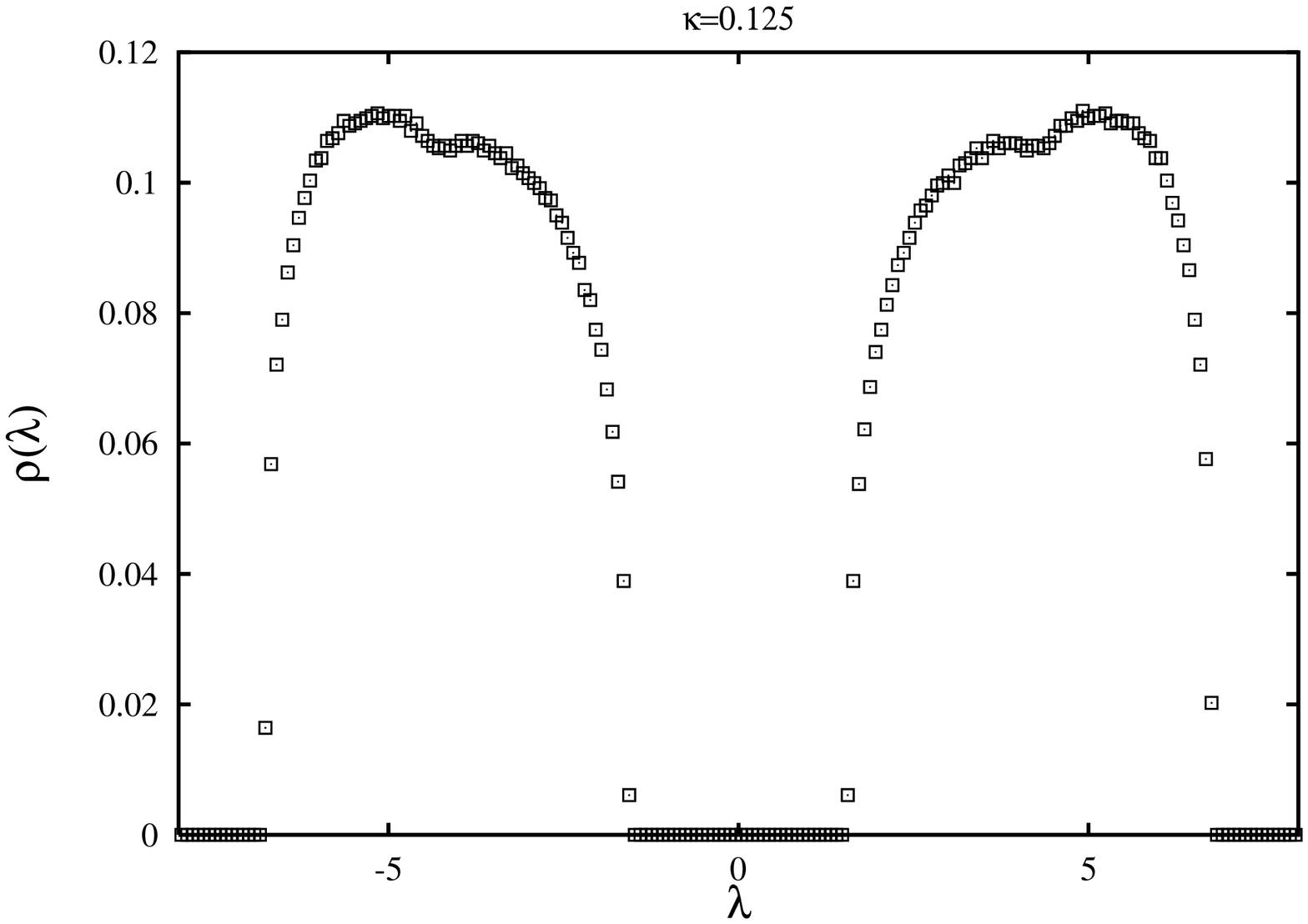}}
\caption{  Spectral densities for the operator $Q_W$ 
for Wilson fermions at $\beta=0$, obtained by 
Kalkreuter [16]. The 
stars (upper), crosses (middle), squares (lower) correspond
to the values of $\kappa$
 equal to $0.25$ ,$0.20$, $0.125$,
 respectively.}
\label{kalk1b}
\end{figure}

Kalkreuter's results are displayed in Fig.~\ref{kalk1b} for $\beta= 4/g^2=0$
(strong coupling) and different values of $\kappa=(2m+8)^{-1}$ 
(Wilson fermions). A similar, although weaker, behaviour is also seen for 
$\beta=1.8$ (weak coupling) and for a range of $\kappa$'s close to the 
critical value of 0.125 \cite{FUKUGITA}.

\begin{figure}[tbp]
\centerline{\epsfxsize=8.6cm \epsfbox{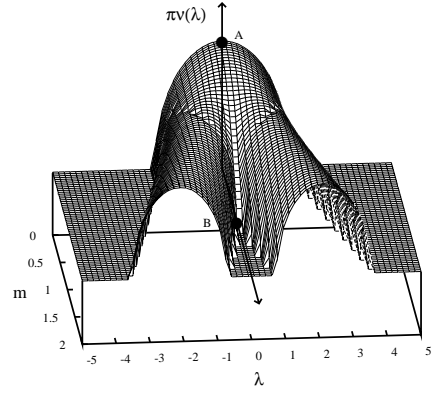}}
\caption{ Spectral function $\pi \nu_Q (\lambda, m)$ (\ref{KK6}) 
 as a function of the  mass $m$ and eigenvalue $\lambda$. 
Point A fixes the normalization for the massless case, point B corresponds to
 the critical mass $m_*=1$. All axes are scaled in units of $\Sigma=1$.
}
\label{kalk1a}
\end{figure}

The behavior of (\ref{KK6}) versus $\lambda$ for different values
$m$ is shown in 
Fig.~\ref{kalk1a}.

In the massless case, (\ref{KK6}) reduces to  (\ref{AR6}) in units of 
$1/\Sigma$, with a half-width equal to $2$. 
In the limit $m>> 2$, the density of states decompose into two pieces,
with $\nu_Q (0 , m) =0$, and a width of the order of $3/\sqrt{2}$ \cite{ZEE},
which is to be contrasted with  4, the size of the semi-circle.
This is a regime, where chiral symmetry is strongly broken by a massive quark. 
For $m_*=1$, the two regions merge into each other. 
In physical units,  the spectral transition occurs for quark masses of the 
order of $m_*\sim -(N/V_4) \left< \overline u u \right>^{-1}$. Typically
$N/V_4\sim 1$ fm$^{-4}$, and for a quark condensate in the range
$(200 MeV)^3$ - $(250 MeV)^3$ we get a critical mass in the range
$m_*\sim 100-200$ MeV, which is rather 
close to the strange quark mass. The spectral transition characterizes the 
transition from a delocalized phase with softly broken chiral symmetry
(Goldstone phase) to a localized phase with strongly broken chiral
symmetry (heavy quark phase). The transition is characterized by the quark 
density of states at zero virtuality, $\nu (\lambda\sim 0, m)$. It is
non-zero in the Goldstone phase, and zero in the heavy-quark phase. This
transition is reminiscent of a Mott-transition from a conductor 
(Goldstone phase) to an insulator (heavy quark phase).

\vskip .5cm
{$\bullet \,\,\,\,$ Staggered Fermions}
\vskip .15cm
For staggered fermions and two colors, it is more appropriate to use the 
Gaussian Symplectic Ensemble (GSE) \cite{JACMANY}, while for three colors 
the GUE. For illustration, consider the GUE for three colors, that is
\be
{\bf Q}_S = \left(\matrix{m & 0\cr 0 & -m \cr}\right) +
\left(\matrix{0 &-iT\cr iT^{\dagger} & 0 \cr}\right)
\label{KK1}
\ee
which is again the sum  of a deterministic $2N\times 2N$ matrix (first
contribution, with $m$ - diagonal block $N\times N$),
and a random hermitean complex $2N\times 2N$ matrix (second 
contribution, with $T$ - block $N\times N$). The resolvent associated to
(\ref{KK1}) is
\be
{\bf G} (z) = \frac z2 \left( 1- 
i\frac{\sqrt{4+m^2-z^2}}{\sqrt{z^2-m^2}}\right) 
\label{KKK1}
\ee
and the corresponding macroscopic spectral distribution is
\be
\nu_Q (\lambda , m ) = 
&&\frac {|\lambda |}{2\pi} \frac{\sqrt{ 4+m^2 - 
\lambda^2}}{\sqrt{\lambda^2-m^2}}\,\,\nonumber\\
&&\times{\Theta }(\lambda^2-m^2)\,\,\Theta ( 4 + m^2 
-\lambda^2 )
\label{KAL3}
\ee
where $\Theta$ are step functions.
The macroscopic spectral distribution is symmetric about the origin, with a 
support from  $m$ to 
$\sqrt{4-m^2}$ to the right, and $-\sqrt{4-m^2}$ and $-m$ to the left. 
The appearance of the gap at zero virtuality is remiscent of the familiar
mass gap in the Dirac equation. Indeed, ${\bf Q}_S$
is nothingh but the Hamiltonian of a quark in a five-dimensional space where 
$\gamma_5$ plays the role of the fifth gamma matrix along the extra (temporal)
direction.

The spectral distribution for the squared staggered operator ${\bf Q}_S^2$,
\be
{\bf Q}_S^2 = \left(\matrix{m^2 + TT^{\dagger} & 0 \cr 0 & m^2 + T^{\dagger} 
T\cr}\right)
\label{KK7}
\ee
that  is
\be
\nu_{Q^2} (\lambda^2 , m ) = \frac{1}{2N} <{\rm Tr}_{2N} (\lambda^2 - 
{\bf Q}_S^2 )>
\label{KK8}
\ee
with the averaging carried out using (\ref{KK2}), can be readily tied to 
(\ref{KAL3}) through
\be
\nu_{Q^2} (\lambda^2 , m ) = \frac 1{|\lambda |} \,\,\nu_Q (\lambda , m )
\label{XKK8}
\ee
This result can also be checked to follow from the large N analysis
using the method suggested by Kazakov or BHZ.

To show this, consider the modified density of states
\be
\tilde{\nu } (\lambda , m ) =  \sqrt{\lambda^2 -m^2}
<< {\rm Tr}_N \delta (\lambda^2-m^2 - T^{\dagger} T ) >>
\label{KAL1}
\ee
which is seen to reduce to (\ref{AR6}) in the massless case, $\tilde{\nu} 
(\lambda , 0 ) = \nu (\lambda )$. Note that the normalization is now to $N$.
In terms of (\ref{KAL1}), the quark condensate reads
\be
\la q^{\dagger} q \ra = -i \int \frac{m}{ |\lambda |} \frac 
1{\sqrt{\lambda^2-m^2}} \,\tilde{\nu} (\lambda , m  )\, d\lambda
\label{KAL2}
\ee
Following Kazakov \cite{KAZAKOV}, we modify 
the Gaussian probability distribution (\ref{KK2}) by adding an 
auxiliary matrix source $A$,
\be
{\bf{P}}_A(T) = \frac{1}{Z_A} e^{-N {\rm Tr}\,(T^{\dagger}T-AT^{\dagger}T)}
\label{modif}
\ee
As a result, the Fourier transform of the resolvent reads
\be
U_A(\tau)=\sqrt{\lambda^2-m^2} \left< \frac{1}{N} 
{\rm Tr} e^{i\tau T^{\dagger}T +i\tau m^2}
 \right>_A
\label{four}
\ee
where the subscript $A$ denotes averaging with the modified probability
distribution (\ref{modif}). In terms of (\ref{four}), the spectral density is
\be
\nu(\lambda , m )=\sqrt{\lambda^2-m^2} 
\int_{-\infty}^{+\infty} \frac{d\tau}{2\pi}
e^{i \tau \lambda^2} U_{A=0}(\tau) 
\label{spec1}
\ee
where $A$ is set to zero, only after the averaging in (\ref{four}) is carried 
out. The result for the macroscopic spectral density is
\be
\tilde{\nu} (\lambda , m  ) = &&\frac 1{2\pi} \sqrt{ 4+m^2 - 
\lambda^2}\,\,\nonumber\\
&&\times{\Theta }(\lambda^2-m^2)\,\,\Theta ( 4 + m^2 
-\lambda^2 )
\label{KAL33}
\ee
in agreement with (\ref{XKK8}-\ref{KAL1}).

The above arguments are subtle at the edge of the spectral distributions 
where  Airy oscillations are expected \cite{BREZIN}. These oscillations are 
at the origin of (\ref{RM3}) in the 
quenched approximation. To analyze them in the staggered case, we need
to magnify the edge points at the level of one level spacing. This can be 
achieved for instance, by taking the microscopic limit 
$N\rightarrow\infty$ with $N\sqrt{\lambda^2-m^2}$ fixed. For the chiral
unitary ensemble, the corresponding microscopic spectral density reads 
(quenched approximation)
\be
\tilde{\nu}_s (\lambda , m ) &=& \frac N2 \sqrt{\lambda^2-m^2}\nonumber \\
& \times& 
\left( J_0^2 (2N \sqrt{\lambda^2-m^2} ) + J_1^2 (2N \sqrt{\lambda^2-m^2} 
)\right)
\label{KAL4}
\ee
In the region $\pm m$, the spectral density dives to zero again, at a rate 
that is comparable to the one discussed in the massless case (\ref{RM3}). 
The effects of the mass
on the unquenched spectral density near zero virtuality are more difficult to 
track down with the methods described above. They may be analyzed by 
Grassmannian techniques as we now discuss.

\vskip .6cm
{\bf 5. \,\,\, Integral Representation : $N_F=1$}\\
 
In a recent analysis, BHZ have put forward an alternative 
method to the Coulomb gas approach and the orthogonal polynomial construction 
to study
 the non-Gaussian character of the density of eigenvalues. Their method 
uses Grassmannian  techniques, and proved to be very elegant to get at the 
oscillating character of the density of states near zero virtuality. Here, we 
will use their method to analyze the elusive case of massive, unquenched  QCD
with three colors. Using the GUE, 
we will discuss the one flavor case in details, and outline the generalization 
to the case of two and more flavors.

The density of eigenvalues $\nu (\lambda , m)$ will be again sought
in terms of the discontinuity of the resolvent $G(z , m)$ along the real
axis. The latter is now given by
\be
{\bf G} (z, m ) = \frac 1{2N} \left< {\rm Tr}_{2N} \left( 
\frac 1{z-{\bf M}}\right) \right>
\label{resolvent}
\ee
with
\be
{\bf M} = \left(\matrix{0& T\cr T^{\dagger} &0\cr}\right)
\label{matrixm}
\ee
The averaging in 
(\ref{resolvent}) is over the distribution ${\bf P} (T)$
for complex $N\times N$ matrices $T$ 
\be
{\bf P} (T) = \frac 1Z e^{- N {\rm Tr} (TT^{\dagger}) }
\,\prod_{F=1}^{N_F}{\rm det}_{2N}\,{\bf M}_F 
\label{weight}
\ee
where $Z$ is an overall normalization, including the fermion determinant.
For each flavor,
\be
{\bf M}_F = \left(\matrix{im & T\cr T^{\dagger} &im \cr}\right)
\label{matrixmf}
\ee

First, let us consider the case of only one flavor $N_F=1$. Following the 
method discussed by BHZ, we can write the unaveraged contribution to the 
resolvent as follows
\be
&&\frac 1{2N} \left( {\rm Tr}_{2N} \, \frac 1{z-{\bf M}} \right) 
\,\,{\rm det}_{2N}\,{\bf M}_1 \nonumber\\
&&=\int \,\,d{\bf \mu} \left( \la a | a \ra + \la b | b \ra \right)
\nonumber\\
&&\times e^{iN z (\la a | a \ra + \la b | b \ra + \la \alpha |\alpha \ra
+\la \beta | \beta \ra )}\nonumber\\
&&\times e^{-iN (\la a | T^{\dagger} | b \ra +\la \alpha |T^{\dagger} |\beta 
\ra - \la \mu | T^{\dagger} |\nu \ra + {\rm h.c.} )}\nonumber\\
&&\times e^{- Nm (\la \mu |\mu \ra + \la \nu | \nu \ra )}
\label{first}
\ee
where $a_i, b_i$ are N-dimensional complex vectors, and 
$\alpha_i,\beta_i,\mu_i,\nu_i$ are N-dimensional Grassmannian vectors.
The measure $d {\bf \mu}$ is graded,
\be
d{\bf \mu} = \prod_{i=1}^N \, [da_i] 
[db_i][d\alpha_i][d\beta_i][d\mu_i][d\nu_i]
\label{measure}
\ee
Averaging (\ref{first}) over $T$ using the Gaussian measure (\ref{weight})
gives,
\be
{\bf G} (z, m) = \frac 1Z \int \prod_{i=1}^N\, d{\bf \mu} &&\,
(\la a | a \ra + \la b | b \ra )\nonumber\\
&&e^{iN z\,(\la a | a \ra + \la b | b \ra + \la \alpha |\alpha \ra
+\la \beta | \beta \ra )}\nonumber\\
&&\times e^{- Nm (\la \mu |\mu \ra + \la \nu | \nu \ra ) -N {\rm Tr_N}}
\label{second}
\ee
with 
\be
{\rm Tr}_N =&&
{\rm Tr}_N( ( -|b\ra\la a| + |\beta\ra\la\alpha| +|\nu \ra \la 
\mu|)\nonumber \\&&
\,\,\,\,\,\,\,\,\,\otimes (- |a\ra \la b| +
 |\alpha \ra \la \beta | + |\mu \ra \la \nu |))
\label{second2}
\ee
The tensor product in (\ref{second}) generates
 four Fermi fields bilinears. They can 
be linearized by introducing four auxiliary complex fields $\sigma_{ij}$ with
$i,j=1,2$, $e.g.$
\be
&&e^{N\la\alpha |\alpha\ra \la \beta|\beta\ra } =\nonumber\\
&&+\frac N{\pi} \int d^2\sigma_{11}e^{-N(\sigma_{11}^* \sigma_{11} 
+\sigma_{11}^* \la\alpha | \alpha \ra + \sigma_{11} \la \beta | \beta \ra)}
\label{aux}
\ee
and so on. In terms of (\ref{second}), the Grassmannian  integrations can be 
undone, and the result is 
\be
{\bf G} (z) \propto \int &&\prod_{i=1}^N [da_i][db_i] [d\sigma] 
(\la a | a\ra + \la b | b \ra )\nonumber\\
&&\times e^{iN z (\la a | a\ra + \la b | b \ra )
- N \la a | a \ra \la b | b \ra }\nonumber \\
&&\times {\rm det} {\bf \Delta}
\label{int0}
\ee
where $4N \times 4N$ block matrix reads
\be
{\bf \Delta} = \left(\matrix {s_1^*& | a \ra \la b|& s_3^* & 0 \cr
|b\ra\la a | & s_1 & 0 & s_4 \cr
s_4^* & 0 &s_2^* & |a\ra\la b |\cr
0& s_3 & |b\ra\la a |& s_2 \cr}\right)
\label{int2}
\ee
with $s_1= iz -\sigma_{11}$, $s_2=m-\sigma_{22}$, $s_3=-\sigma_{12}$
and $s_4=-\sigma_{21}$. The 
matrix (\ref{int2}) is sparse. It can be rotated to 
an $N$ block-diagonal form of $4\times 4$ matrices. Hence
\be
{\rm det} {\bf \Delta} = &&
( (a^2 b^2)^2 -a^2b^2 \sum_{i=1}^4 s_is_i^* + 
|s_1s_2-s_3s_4|^2))\nonumber\\
&&\times \left(|s_1s_2-s_3s_4|^2\right)^{N-1}
\label{det1}
\ee
with $a^2 =\la a | a \ra$ and $b^2= \la b | b \ra$. 
The integrals in (\ref{int2}) may be simplified by introducing the new 
variables
\bea
s_{ij}&=&\sigma_{ij}-iz \delta_{11} - m\delta_{22},\\ 
s_{ij}^*&=&\sigma_{ij}^*-iz\delta_{11} - m\delta_{22}.
\eea
and $x=a^2$ and $y=b^2$. $\delta_{11}$ is non-zero for $i=j=1$ and similarly 
for $\delta_{22}$. 
Thanks to the Gaussian term in the integration over $\sigma$ and $z=\lambda 
+i\epsilon$, we can freely 
shift the contour of integration, in particular we can treat matrices
${\bf s}$ and ${\bf s^*}$ as complex conjugate. Hence
\bea
G(z, m) &=& {\cal N} e^{-Nm^2}\int_0^{\infty}dx~dy~ds_{ij}~ds_{ij}^* 
\nonumber\\
&\times&(x+y)(xy)^{N-1}
e^{-Nxy-N\tr s s^{\dagger} + N m (s_{22} + s_{22}^*)}\nonumber\\
&\times& e^{i \zeta (x+y-s_{11}-s_{11}^*)}
	e^{\zeta^2/N}
	\det \Delta( s s^{\dagger},xy)
\label{g1}
\eea
where ${\cal N}$ is a normalization constant and
\beq
\det {\bf \Delta} = (x^2y^2 - xy \tr {\bf s\, s}^{\dagger} +
 \det {\bf s\, s}^{\dagger})
	(\det {\bf s\, s}^{\dagger})^{N-1}.
\label{det2}
\eeq
Notice that the saddle point in $s_{ij}$ has a symmetry with respect to
the unitary 
rotations of ${\bf s}$ for $m=0$. Only the ${\cal O}(1)$ term breaks this 
symmetry. The relations 
(\ref{g1},\ref{det2}) constitute the integral representation for 
the resolvent ${\bf G}(z,m)$ in the complex z-plane, for one massive flavor
and finite dimension $N$. Its singularity along the real axis 
$z=\lambda +i\epsilon$ corresponds to
the distribution of eigenvalues of the QCD Dirac operator with one massive
flavor as modeled by the chiral random matrix ensemble. 

In the thermodynamical limit $N\rightarrow\infty$, the integral in 
(\ref{g1}) is dominated by the saddle point configurations. The latter are 
the extrema of 
\be
{\bf S}_{\rm eff} = &&xy -{\rm ln} \,\,xy -iz (x+y) \nonumber\\
&&+{\rm Tr} ({\bf s\, s}^{\dagger} )-
{\rm ln} ( {\rm det}{\bf s\, s}^{\dagger} )
\label{action}
\ee
The extremum of (\ref{action}) coincides with the extremum in the quenched 
approximation (BHZ), leading to a semi-circular distribution of eigenvalues. 
The fermion determinant does not affect the eigenvalue distributions in the
large $N$ limit. As observed in \cite{JAC} and discussed by BHZ, the 
saddle point approximation breaks down at $\lambda =0$ and also the edges of 
the semi-circular distribution due to Airy-type singularities (quenched)
and zero modes (unquenched).

\vskip .6cm
{\bf 6.\,\,\, Microscopic Limit : Massless Case } \\

To investigate the behavior of the spectral density near the 
origin for the GUE, we will use the microscopic limit discussed in \cite{JAC}.
 We will simply
blow up the spectrum near $\lambda =0$ over a scale of the order of one-level 
spacing. This is achieved by considering $Nz=\zeta={\cal O} (1)$ in the region 
$z=0$ (zero virtuality), and track down the $1/N$ corrections. Since the mass 
term and the $1/N$ corrections violate the saddle point invariance under 
unitary rotations, care is to be used to disentangle the soft directions
from the hard directions. All soft directions have 
to be integrated out exactly,
 for an accurate estimate of the saddle around $z=0$.

First, let us consider the massless case $m=0$, where only the ${\cal O}(1)$
terms break the unitary symmetry of ${\bf s}$. To make the latter manifest,
we use the parameterization
as
\bea
{\left(\matrix{s_{11} & s_{12}\cr s_{21}&s_{22}\cr}\right)} &&=
{\left(\matrix{\sqrt{u} & 0 \cr \sqrt{v}e^{i\beta} & \sqrt{w}e^{i\gamma} \cr}
\right)}
{\left(\matrix{\cos\theta & \sin\theta \cr -\sin\theta & \cos\theta
\cr}\right)}\nonumber \\ &&\times
\left(\matrix{e^{i\phi} & 0 \cr 0 & e^{i\psi}\cr}\right).
\eea
so that
\bea
\tr s s^{\dagger} &=& u+v+w,\\
\det s s^{\dagger} &=& u w\\
s_{11}+s_{11}^* &=& 2 \sqrt{u} \cos\theta \cos\phi.
\eea
Following BHZ let us introduce also
\bea
xy &=& p,\\ 
x+y &=& 2\sqrt{p}\cosh \omega.
\eea
With these notations we have 
neglecting the ${\cal O}(1/N)$ term in the exponent
\bea
G(z, 0)&=& {\cal N}^{'}  
\int_0^{\infty}\cosh\omega d\omega\int_0^{\infty}\frac{dp}
{\sqrt{p}}\nonumber\\&\times&
\int_0^{\infty}u du dv dw \int_0^{2\pi}d\phi d\psi d\beta d\gamma \\
\no
&\times&
\int_0^{\pi/2}\sin\theta \cos\theta d\theta  
   p^N (uw)^{N-1}\nonumber\\&\times&(p^2-p(u+w+v)+uw) \\ 
\no
& \times & e^{-N(u+w+v+p)}\nonumber\\ &\times&
e^{2i\zeta\sqrt{p}\cosh\omega}e^{2i\zeta\sqrt{u}\cos\phi\cos\theta}.
\eea
The $\psi,\beta$ and $\gamma$ angular integrals can be trivially performed.
Also the $v$ integral is trivial.
The $\theta$ and $\phi$ integrals can be performed using
\bea
&&\int_0^{2\pi}d\phi\int_0^{\pi/2}d\theta~\sin\theta~\cos\theta 
e^{-i\zeta\sqrt{u}\cos\theta\cos\phi}\nonumber\\
=&& 2 \pi \frac{J_1(2\zeta\sqrt{u})}
{2\zeta\sqrt{u}}
\eea 
and the $\omega$ integral by
\bea
\int_0^{\infty}d\omega\cosh\omega e^{2i\zeta\sqrt{p}\cosh\omega}
&=& K_1(-2i\zeta\sqrt{p})\\ 
\no
&=& -\frac{\pi}{2}(J_1(2\zeta\sqrt{p})+iN_1(2\zeta\sqrt{p})).
\eea
For the imaginary part of $G(\zeta ,0)$ we get
\bea
\rho(\zeta/N , 0)&=&-\pi^2 \sqrt{2}\left( \frac{N}{\pi} \right)^{7/2}
e^{3N}
\int_0^{\infty}\frac{dp}{\sqrt{p}}
du\frac{dw}{w}\nonumber\\
&\times& (p^2 - p(u+w+\frac{1}{N})+uw) \nonumber \\
& \times & (puw)^Ne^{-N(p+u+w)}
 J_1(2\zeta\sqrt{p})\frac{J_1(2\zeta\sqrt{u})}{2\zeta\sqrt{u}}.
\eea 
Where we have explicitly wrote down the normalization
(we used Stirling  formula to rewrite the normalization in this form).
This integral is dominated by the neighborhood of the saddle point
\beq
p_c = u_c = w_c =1.
\eeq
Writing
\bea
p &=& 1 + \frac{p'}{\sqrt{N}},\\ \no
u &=& 1 + \frac{u'}{\sqrt{N}},\\ \no
w &=& 1 + \frac{w'}{\sqrt{N}}, \label{here}
\eea
we have
\bea
&&p^2 -p(u+w+1/N)+uw = \nonumber \\&&
\frac{1}{N}((p')^2-1 -
u'p'+w'(u'-p')-\frac{1}{\sqrt{N}}p').
\eea
The leading contribution coming from $(p')^2-1$ cancels out. We are
therefore led to consider the  next order terms. We observe that due
to the symmetry in $p'$ and $u'$ the terms containing $w'$ cancel.
Using (\ref{here}) we have
\bea
&&\frac{J_1(2\zeta\sqrt{u})}{\sqrt{u}}\nonumber \\
=&&J_1(2\zeta)-\frac{u'}{\sqrt{N}}
\zeta J_2(2\zeta)+\frac{(u')^2}{2N}\zeta^2 J_3(2\zeta) + \dots \nonumber\\  
&&= F_0 + \frac{u'}{\sqrt{N}}F_1+\frac{(u')^2}{N}F_2 + \dots 
\eea
\bea
&&e^{-N(u-\log u)}\frac{J_1(2\zeta\sqrt{u})}{\sqrt{u}}\nonumber\\=&&
F_0 + \frac{1}{\sqrt{N}}\left(\frac{(u')^3}{3}F_0 + u' F_1\right) \nonumber\\
&&+ \frac{1}{N}\left( (\frac{(u')^6}{18}-\frac{(u')^4}{4})F_0
+\frac{(u')^4}{3}F_1 + (u')^2 F_2 \right).
\eea
Using this expansion we get
\bea
\rho(\zeta/N) & = -2&\zeta J_1(2\zeta) J_3(2\zeta) +
J_1(2\zeta)J_2(2\zeta)+\zeta J_2^2(2\zeta) \nonumber \\&=&
2\zeta\left(J_1^2(2\zeta)-J_0(2\zeta)J_2(2\zeta)\right).
\label{RM3again}
\eea
in agreement with (\ref{RM3}) for $N_F=1$.

\vskip .6cm
{\bf 7.\,\,\, Microscopic Limit : Massive Case }\\
 
Let us consider the general case of one massive flavor.
%
The inclusion of the mass corresponds to analyzing (\ref{g1}).
Three different regimes may be be considered, depending on the strength of the 
sea quark mass. The regime $Nm >>1$, corresponds to 
the case where the sea mass is comparable to the valence mass, in the physical 
(chiral) limit. In this case, the mass term in (\ref{g1}) contributes to the 
saddle point equations along $s$ and $s^{\dagger}$. The saddle point equations 
in $x$ and $y$ remain unaffected, and the resulting spectral distribution in 
the large $N$ limit is a semi-circle. This result is in agreement with the 
Coulomb gas analysis. We expect, the microscopic spectral density to be 
unaffected by the mass effect. In the  regime $Nm<< 1$, the mass effects are 
sub-leading compared to the ${\cal O} (1)$ terms. This regime 
corresponds to the massless case 
discussed above. Both the macroscopic and microscopic spectral densities are
expected to be unaffected by such masses. 
The most interesting regime, is the one for which $Nm\sim 1$, as we now 
discuss.

In the case $\mu = Nm = {\cal O}(1)$,
\bea
G(z,m) &=&  {\cal N}^{'} \int_0^{\infty}dx~dy~ds_{ij}~ds_{ij}^* 
(x+y)(xy)^{N-1}\\ \no
     &\times&e^{-Nxy-N\tr s s^{\dagger}}
e^{i \zeta (x+y-s_{11}-s_{11}^*)}e^{-\mu(s_{22}+s_{22}^*)}\\ \no
&\times& e^{(\zeta)^2/N}
	\det {\bf \Delta} ( {\bf s\, s}^{\dagger},xy).
\label{g2}
\eea
We can use the same parameterization as before and observe that
\beq
s_{22}+s_{22}^*=2\sqrt{v}\sin\theta \cos(\beta+\psi)
+2\sqrt{w}\cos\theta \cos(\gamma + \psi).
\eeq
Let us introduce
\beq
t = \cos\theta.
\eeq
We have
\bea
&&G(z,m)= {\cal N}^{'}  
\int_0^{\infty}\cosh\omega d\omega\int_0^{\infty}\frac{dp}
{\sqrt{p}} \\ \nonumber  &\times& 
\int_0^{\infty}u du dv dw \int_0^{2\pi}d\phi d\psi d\beta d\gamma \\
\no
&\times&
\int_0^1 t~dt 
   p^N (uw)^{N-1}(p^2-p(u+w+v)+uw) \\ 
\no
& \times & e^{-N(u+w+v+p)}
e^{2i\zeta \sqrt{p}\cosh\omega}e^{2i\zeta\sqrt{u}t\cos\phi}\\ \no
&\times &
e^{-\mu\sqrt{w}t\cos\gamma -\mu\sqrt{v}\sqrt{1-t^2}\cos\beta}.
\eea
We perform the integrals over $\omega$ (as before) and over angular
variables $\phi$, $\psi$, $\beta$ and $\gamma$ using
\bea
\int_0^{2\pi} d\phi~e^{ix\cos\phi}&=&2\pi J_0(x),\\
\int_0^{2\pi} d\phi~e^{-x\cos\phi}&=&2\pi I_0(x).
\eea
We get
\bea
&&\rho(\zeta /N,\mu/N)= -\pi^2\sqrt{2}\left( \frac{N}{\pi} \right)^{7/2}
e^{3N}
\int_0^{\infty}dp~dv~
du\frac{dw}{w}\nonumber \\
&& \int_0^1 t~dt(p^2 - p(u+w+v)+uw) \\ \no
& \times & (puw)^Ne^{-N(p+u+w+v)}
\frac{J_1(2\zeta \sqrt{p})}{\sqrt{p}}J_0(2\zeta t\sqrt{u})\\ \no
&\times&
I_0(2\mu\sqrt{1-t^2}\sqrt{v}) I_0(2\mu t\sqrt{w}).
\label{exact}
\eea 
This is an explicit 
integral representation of the spectral distribution in the 
massive case and for finite $N$. No
 approximation has been used in getting from 
(\ref{g2}) to (\ref{exact}). We observe
that the integral over $v$ can be performed in the
leading order and that to this order we can replace
$I_0(2\mu\sqrt{1-t^2}\sqrt{v})$ by one.
Using this fact and the formula:
\beq
\int_0^1 t~dt J_0(xt)I_0(yt)=\frac{y I_1(y)J_0(x)+x
J_1(x)I_0(y)}{x^2+y^2}
\eeq
we have 
\bea
&&\rho(\zeta ,\mu) \propto
\int_0^{\infty}dp~
du\frac{dw}{w}(p^2 - p(u+w+\frac{1}{N})+uw)\\ \no &\times& 
 (puw)^Ne^{-N(p+u+w)}
\frac{J_1(2\zeta \sqrt{p})}{\sqrt{p}} \\ \no &\times&
\frac{\mu\sqrt{w}I_1(2\mu\sqrt{w})J_0(2\zeta \sqrt{u})
+\zeta \sqrt{u}J_1(2\zeta \sqrt{u})I_0(2\mu\sqrt{w})}{2((\zeta )^2u
+\mu^2 w)}.
\eea
The integral is dominated again by the neighborhood of the saddle point
at $p_c=u_c=w_c=1$. We need as before terms ${\cal O}(1/N)$ to calculate
it.

Expanding the above formula 
as before, we obtain after somewhat lengthy algebra,
\bea
\rho(\zeta,\mu)&=&
2 I_0 \left( J_0^2\lambda\cos\tau - J_0 J_1 \frac{3+\cos 4\tau}{4}\right.
 \nonumber \\
 &-&  \left. J_1^2\frac{-8 \lambda^2\cos\tau+\cos 3\tau-\cos
5\tau}{8\lambda}\right) \nonumber \\
&-&2 I_1\sin\tau\left(J_0^2\cos\tau+J_0 J_1\frac{\cos 2\tau
\sin^2\tau}{\lambda}\right. \nonumber \\
 &+& \left. 2 J_1^2\cos\tau\sin^2\tau \right).
\label{FINAL}
\eea
with the parameters
\bea
\zeta &=& \lambda \cos \tau,\\ \no
\mu   &=& \lambda \sin \tau.
\eea
The Bessel functions in (\ref{FINAL}) are defined as $J_n = J_n (2\zeta )$
and $I_n = I_n (2\mu )$. For $\tau=0$, we recover (\ref{RM3}) for $N_F=1$
and (\ref{RM3again}).

Clearly, sea masses of the order $Nm = {\cal O} (1)$ do affect the character
of the spectral oscillations near zero virtuality. This in turn would imply, 
new sum rules for the moment of the eigenvalues of the massive QCD Dirac 
operator. The very non-local character of the fermion determinant upsets
the universality at zero virtuality.

The present analysis can be extended to arbitrary flavors $N_F =2,3, ...$.
Indeed, for $N_F$ flavors the bosonization of the fermion bilinears requires 
the introduction of $2(N_F+1)^2$ auxiliary fields, which we label generically
by the matrices ${\bf s}$ and ${\bf s}^{\dagger}$. After 
integration, the analog of the 
determinant ${\bf \Delta}$ in (\ref{int2}) splits into 
$(2N_F+2)\times(2N_F+2)$ blocks each of size $N \times N$. By analogy with the 
one flavor case, the determinant can be reduced to the product of $N$ 
determinants of size $(2N_F+2)\times(2N_F+2)$. The result, can be generically 
written in the form
\bea
{\bf \Delta}_{N_F} = P_{N_F+1}( (a^2 b^2), {\bf ss}^{\dagger}) {\rm det}^{N-1}
({\bf s s}^{\dagger})
\eea
where $P$ is a polynomial
of degree $N_F+1$ in $a^2b^2$ with coefficients depending on the invariants
of the combination ${\bf ss}^{\dagger}$. So the 
number of flavors enters explicitly
as the degree of the polynomial and implicitly as the dimension of the 
matrix ${\bf s}$. In the large $N$ limit the integral is dominated by the 
saddle point, in which the polynomial does not contribute. 
Hence, the saddle point configurations are the minima of the action 
(\ref{action}), resulting into a semi-circular distribution of eigenvalues.
This result is totally in agreement with the Coulomb gas argument. We expect, 
the above arguments near zero virtuality to carry (tediously) to the massive 
case.

\vskip .6cm
{\bf 8.\,\,\, Conclusions} \\

Using chiral random matrix models as inspired by QCD spin and 
flavor symmetries, we have discussed 
the effects of a light quark mass on the QCD Dirac spectrum. In the 
thermodynamical limit (large N), the quarks play a subleading role on the 
macroscopic spectral density, except near zero virtuality. For one
flavor and three colors (GUE), we have shown that the microscopic spectral 
distribution is affected 
by quark masses $Nm ={\cal O} (1)$. The non-local character of the fermion 
determinant causes the spectral oscillations to be flavor and mass dependent
near zero virtuality. New sum rules of the type discussed by Leutwyler and 
Smilga are therefore expected. Our arguments extend to more than one flavor. 

Kalkreuter's spectral distributions from lattice SU(2) gauge theories make use
of a hermitean variant of the Dirac operator. In the quenched approximation
and for Wilson fermions,
they can be understood using a Gaussian Orthogonal matrix model,
reminiscent of the one used in quantum Hall fluids. In contrast to the 
preceding discussion, here the quark mass survives the thermodynamical limit. 
For one flavor,
Kalkreuter's spectral  distributions show that a spectral transition occurs 
for quarks masses of the order of the strange quark mass. The transition is 
reminiscent of the Mott-transition from metals to insulators, with the density 
of states at zero virtuality playing the role of an order parameter. 

The similarity between the spectral distribution generated from lattice 
simulations and the one obtained from the chiral random matrix model with
massive quarks suggests that the lattice results may be robust to cooling.
If that is the case, then we suggest that a properly regulated quark 
condensate in the presence of finite current quark masses can be obtained from
cooled lattice configurations, in agreement with the general lore of random 
matrix theory. This point is worth checking on the lattice.

The behavior of the spectral density near zero virtuality and for massive 
quarks, is of relevance to finite temperature lattice simulations of QCD 
\cite{JAC3}. The 
sensitivity of the lattice results both to the masses and finite volumes
(whether temporal or spatial) in the infrared regime,
are important for a reliable assessment of the chiral condensate, and hence 
the study of the spontaneous breaking of chiral symmetry in finite systems
both in vacuum and matter. 
The role of the current quark masses and the number of flavors
in the QCD phase transition is of paramount importance for issues such as 
the order and character of the transition. Some of these issues will be 
discussed elsewhere.

\vglue 0.6cm
{\bf \noindent  Acknowledgements \hfil}
\vglue 0.4cm
 
We are grateful to  Dr Thomas Kalkreuter for sending us his data.
 M.A.N. wishes to thank  
G\'{a}bor Papp, for discussions and help with computer graphics.
This work was supported in part  by the US DOE grant DE-FG-88ER40388
and by the Polish Government Project (KBN)  grant  2P03B19609.

\vskip 1cm
\setlength{\baselineskip}{15pt}

\end{document}